\documentclass[nofootinbib,twocolumn,superscriptaddress,showpacs,preprintnumbers,amsmath,amssymb,aps,prd]{revtex4}
\usepackage{amssymb} \usepackage{graphicx} \usepackage{color} % $Id: defs.tex,v 1.4 2009/01/23 14:29:23 pablo Exp $

\def\elpar#1{$\lbrace \tilde g_{ij},\,\widetilde K_{ij}\rbrace$#1}
\newcommand{\Long}{\mathbb{L}}

\newcommand{\fmin}{{f_{\mathrm{min}}}}
\newcommand{\fmax}{{f_{\mathrm{max}}}}

\def\ltsima{$\; \buildrel < \over \sim \;$}
\def\ltsim{\lower.5ex\hbox{\ltsima}}
\def\gtsima{$\; \buildrel > \over \sim \;$}
\def\gtsim{\lower.5ex\hbox{\gtsima}}

\def\newacronym#1#2#3{\gdef#1{#3 (#2)\gdef#1{#2}}}

\newacronym{\NSF}{NSF}{National Science Foundation}
\newacronym{\NASA}{NASA}{National Aeronautics and Space Administration}
\newacronym{\lisa}{LISA}{the Laser Interferometer Space Antenna}
\newacronym{\ligo}{LIGO}{Laser Interferometer Gravitational-wave Observatory} 
\newacronym{\Caltech}{Caltech}{California Institute of Technology}
\newacronym{\MIT}{MIT}{Massachusetts Institute of Technology}
\newacronym{\sph}{SPH}{smooth particle hydrodynamics}
\newacronym{\tsi}{TSI}{the Terascale Supernova Initiative}
\newacronym{\wmap}{WMAP}{the Wilkinson Microwave Anisotropy Probe}
\newacronym{\decigo}{DECIGO}{the Deci-Hertz Interferometric Gravitational-wave Observatory} 
\newacronym{\cmbr}{CMBR}{cosmic microwave background}
\newacronym{\ibbh}{IBBH}{intermediate binary black hole}
\newacronym{\bdj}{BDJ}{Brans-Dicke-Jordan}
\newacronym{\bbo}{BBO}{Big Bang Observer}
\newacronym{\decigo}{DECIGO}{Deci-Hertz Gravitational-Wave Observatory}

\def\MPR#1{{\it Moving Puncture Recipe}#1 (MPR#1)\gdef\MPR{MPR}}
\def\ahz#1{apparent horizon#1 (AH#1)\gdef\ahz{AH}}
\def\CM#1{center-of-mass#1 (CM#1)\gdef\CM{CM}}
\def\CLA#1{close-limit approximation#1 (CLA#1)\gdef\CLA{CLA}}
\def\pnw#1{post-Newtonian#1 (PN#1)\gdef\pnw{PN}}
\def\nr#1{numerical relativity#1 (NR#1)\gdef\nr{NR}}
\def\qnm#1{quasi-normal mode#1 (QNM#1)\gdef\qnm{QNM}}
\def\isco#1{innermost stable circular orbit#1 (ISCO#1)\gdef\isco{ISCO}}
\def\eos#1{equation of state#1 (EOS#1)\gdef\eos{EOS}}
\def\tov#1{Tolman-Oppenheimer-Volkoff#1 (TOV#1)\gdef\tov{TOV}}
\def\ns#1{neutron star#1 (NS#1)\gdef\ns{NS}}
\def\bbh#1{binary black hole#1 (BBH#1)\gdef\bbh{BBH}}
\def\bhns#1{black hole -- neutron star#1 (BHNS#1)\gdef\bhns{BHNS}}
\def\nsns#1{neutron star -- neutron star#1 (NSNS#1)\gdef\nsns{NSNS}}
\def\emri#1{extreme mass-ratio inspiral#1 (EMRI#1)\gdef\emri{EMRI}}
\def\emrb#1{extreme mass-ratio binaries#1 (EMRB#1)\gdef\emrb{EMRB}} 
\def\grb#1{gamma-ray burst#1 (GRB#1)\gdef\grb{GRB}}
\def\imbh#1{intermediate mass black hole#1 (IMBH#1)\gdef\imbh{IMBH}}
\def\smbh#1{supermassive black hole#1 (SMBH#1)\gdef\smbh{SMBH}}
\def\bh#1{black hole#1 (BH#1)\gdef\bh{BH}}
\def\ulx#1{ultra-luminous x-ray source#1 (ULX#1)\gdef\ulx{ULX}}
\def\nps#1{Newman-Penrose#1 (NP#1)\gdef\nps{NP}} 
\def\lmxbs{low-mass x-ray Binaries (LMXBs)\gdef\lmxbs{LMXBs}\gdef\lmxb{LMXB}} 
\def\lmxb{low-mass x-ray Binary (LMXB)\gdef\lmxbs{LMXBs}\gdef\lmxb{LMXB}} 
\def\cv#1{constraint violation#1 (CV#1)\gdef\cv{CV}}
\def\cs#1{constraint satisfaying#1 (CS#1)\gdef\cs{CS}}

\newcommand\apjl{\ref@jnl{ApJ}}%
\newcommand\mnras{\ref@jnl{MNRAS}}%

\begin{document}

\title{Binary black hole evolutions of approximate puncture initial data}

\author{Tanja Bode}
\affiliation{Center for Gravitational Wave Physics\\
The Pennsylvania State University, University Park, PA 16802}
\author{Pablo Laguna}
\affiliation{Center for Relativistic Astrophysics and
School of Physics\\
Georgia Institute of Technology, Atlanta, GA 30332}
\author{Deirdre M. Shoemaker}
\affiliation{Center for Relativistic Astrophysics and
School of Physics\\
Georgia Institute of Technology, Atlanta, GA 30332}
\author{Ian Hinder}
\affiliation{Max-Planck-Institut f\"ur Gravitationsphysik,
  Albert-Einstein-Institut, Golm, Germany}
\author{Frank Herrmann}
\affiliation{Center for Scientific Computation and Mathematical Modeling, University of Maryland, College
  Park, MD 20742}
\author{Birjoo Vaishnav} 
\affiliation{Department of Physics and Astronomy\\
University of Texas At Brownsville, Brownsville, TX 78520}

%%%%%%%%%%%%%%%%%%%%%%%%%%%%%%%%%%%%%%%%%%%%%%%%%%%%%%%%%%%%%%%%%%%%%%%%%%%%
\begin{abstract} 
  Approximate solutions to the Einstein field equations are a valuable
  tool to investigate gravitational phenomena. An important aspect of
  any approximation is to investigate and quantify its regime of
  validity. We present a study that evaluates the effects that
  approximate puncture initial data, based on \emph{skeleton}
  solutions to the Einstein constraints as proposed by
  \citet{2004PhRvD..69l4029F}, have on numerical evolutions.  Using
  data analysis tools, we assess the effectiveness of these
  constraint-violating initial data and show that the matches of
  waveforms from skeleton data with the corresponding waveforms from
  constraint-satisfying initial data are $\gtrsim 0.97$ when the total
  mass of the binary is $\gtrsim 40M_\odot$. In addition, we
  demonstrate that the differences between the skeleton and the
  constraint-satisfying initial data evolutions, and thus waveforms,
  are due to negative Hamiltonian constraint violations present in the
  skeleton initial data located in the vicinity of the
  punctures. During the evolution, the skeleton data
  develops both Hamiltonian and momentum constraint violations that
  decay with time, with the binary system relaxing to a
  constraint-satisfying solution with black holes of smaller mass and
  thus different dynamics.
\end{abstract}

\maketitle
%%%%%%%%%%%%%%%%%%%%%%%%%%%%%%%%%%%%%%%%%%%%%%%%%%%%%%%%%%%%%%%%%%%%%%%%%%%%%
\section{Introduction}
\label{sec:intro}

With the developments of the past few years, numerical relativity
simulations of \bbh{} systems from inspiral to merger are now
feasible, almost routine. Most importantly, they are quickly becoming
a potent tool to study highly relevant astrophysical
phenomena. Approximations such as those provided by \pnw{} theory have
also proven to be valuable tools.  They have the appeal of avoiding
the computational complexities associated with finding exact solutions
to the Einstein field equations.  As the demand for more efficient
simulations increases, it is desirable to consider approximate
methodologies in conjunction with numerical relativity approaches. A
natural ``marriage'' in this regard, which is the focus of this work,
is to consider full Einstein evolutions of approximately
constraint-satisfying initial data.

In general relativity, constructing initial data requires solving the
Einstein constraints, a coupled set of elliptic equations
(see~\citet{Baumgarte:2002jm} for a review on the mathematical
foundations of numerical relativity and~\citet{Cook:2000vr} for
constructing initial data). Thus, in general obtaining solutions to the Einstein
constraints necessitates solving elliptic equations, which is a complex
numerical problem. When \bh{} excision is
used, the solvers are
non-trivial~\cite{Thornburg2000:multiple-patch-evolution,Pfeiffer:2002wt,Pfeiffer:2002xz}
because of the excision boundaries.  Even without excision, developing
constraint solvers is demanding~\cite{Ansorg:2004ds} and often
requires introducing simplifying assumptions such as spatial conformal
flatness.

Flexibility is also a very important issue.  The family of problems
addressed by numerical relativity is quickly expanding, involving
non-traditional \bh{} systems beyond the two-body problem~\cite{Campanelli:2007ea,Lousto:2007rj}. Without
modifications to the standard initial data methodology, there will be
limitations on the class of problems one is able to consider.

The focus of the present work is on the full Einstein numerical
evolution of constraint-violating or approximate initial
data. Evolutions of constraint-violating \bbh{} initial data have been
considered in the past. They were mostly done in the context of
superposed Kerr-Schild
\bh{s}~\cite{Sperhake2005a,Brandt00,Marronetti00a,Matzner98a}. More
recently, constraint-violating initial data for punctures has been
used for multiple \bh{}
evolutions~\cite{Campanelli:2007ea,Lousto:2007rj}.

The difference with previous studies is in the building blocks used to
construct the data.  In
Refs.~\cite{Dennison:2006nq,Campanelli:2007ea,Lousto:2007rj}, the
initial data sets were built from perturbative solutions of single
punctures (boosted and/or spinning). Our approach, on the other hand,
follows closely the \emph{skeleton} solutions of the Einstein
equations introduced by \citet{2004PhRvD..69l4029F}. These solutions
are derived from the full Arnowitt-Deser-Misner (ADM) Hamiltonian with
the \bh{s} represented by point-like sources modeled by Dirac delta
function distributions. We consider configurations of non-spinning,
equal-mass \bbh{s} in quasi-circular orbits and investigate how well
the evolution of these initial data is able to reproduce the
corresponding results of constraint-satisfying initial data.  We
assess the effectiveness of the skeleton initial data by computing the
matches with waveforms from constraint-satisfying initial data
evolutions. We find that the differences in the evolutions, and thus
waveforms, are due to negative Hamiltonian constraint violations
present in the skeleton initial data.  We observe that, during the
course of the evolution, the skeleton data develops both Hamiltonian
and momentum constraint violations which both propagate away and decay 
over time while the binary system relaxes to a constraint-satisfying 
solution with \bh{s} of smaller mass and thus different dynamics.

In Sec.~\ref{sec:ID}, we derive the procedure for constructing
skeleton puncture initial data. In Sec.~\ref{sec:quasi}, we focus on
quasi-circular configurations of equal-mass, non-spinning \bbh{s},
and, using the effective potential method~\cite{Cook:2000vr}, we
compare binding energies between skeleton and corresponding
constraint-satisfying initial data. In Sec.~\ref{sec:negcvs}, we
investigate the structure of the Hamiltonian constraint violations in
the skeleton data. In Sec.~\ref{sec:evolutions}, we present results of
the evolutions. Sec.~\ref{sec:single_punc} presents an analysis of the
nature of the constraint violations with a model involving a single
puncture. In Sec.~\ref{sec:dataanalysis}, we discuss the impact of
using waveforms from skeleton evolutions on data analysis.
Conclusions are given in Sec.~\ref{sec:conclusions}.

The numerical simulations and results were obtained with the MayaKranc
infrastructure as described
in Refs.~\cite{Herrmann:2006ks,2007ApJ...661..430H,vaishnav-2007,2007arXiv0706.2541H}.

%%%%%%%%%%%%%%%%%%%%%%%%%%%%%%%%%%%%%%%%%%%%%%%%%%%%%%%%%%%%%%%%%%%%%%%%%%%%%
\section{Skeleton Initial Data}
\label{sec:ID}

The traditional approach to constructing initial data in numerical
relativity involves specifying the pair \elpar{,} where $\tilde
g_{ij}$ is the intrinsic 3-metric to a $t$ = constant hypersurface
$\Sigma_t$, and $\widetilde K_{ij}$ denotes its extrinsic
curvature. We use the index convention that Latin indices in the first
part of the alphabet denote 4-dimensional spacetime indices and those
from the middle denote 3-dimensional spatial indices. The pair
\elpar{} must satisfy the Einstein constraint equations:
\begin{eqnarray}
  \widetilde R + \widetilde K^2 - \widetilde K_{ij} \widetilde K^{ij} 
  &=& 16\,\pi\,\tilde\rho  \label{hamcons}\\
  \widetilde\nabla_j \widetilde K^{ij} - \widetilde\nabla^i \widetilde K 
  & =& 8\,\pi\,\tilde j^i \, .\label{momcons}
\end{eqnarray}
Equations (\ref{hamcons}) and (\ref{momcons}) are respectively known as the
Hamiltonian and momentum constraints.  The
operator $\widetilde\nabla_i$ denotes covariant differentiation with
respect to $\tilde g_{ij}$ and $\tilde R_{ij}$ its associated Ricci
tensor. We follow the notation $\tilde K\equiv\tilde g^{ij}\tilde
K_{ij}$ and $\tilde R\equiv\tilde g^{ij}\tilde R_{ij}$.

Although we are interested in vacuum spacetimes of \bh{} systems, we
have kept the matter sources $\tilde\rho$ (total energy density) and
$\tilde j^i$ (momentum density).  This is so we are able, as in
Ref.~\cite{2004PhRvD..69l4029F}, to represent the \bh{s} as point-like
sources modeled with Dirac delta distributions.

The constraints Eqs.~(\ref{hamcons}) and (\ref{momcons}) yield four
equations; there are, thus, eight freely specifiable pieces in the
data \elpar{.}  These free data can be used to single out the physical
system under consideration (e.g. orbiting binary \bh{s}) as well
as to simplify solving the Einstein constraints. An elegant approach
to identify the four pieces in \elpar{} that are fixed from
solutions to the constraints was given in \citet{1979sgrr.conf...83Y},
based on work by \citet{Lichnerowicz44} and others. The method is
based on the following conformal transformations and tensorial
decompositions:
\begin{eqnarray}
\label{eq:gconf}
\tilde g_{ij} &=& \psi^4\,g_{ij} \\
\label{eq:Kdecomp}
\widetilde K_{ij} &=& \widetilde A_{ij}+\frac{1}{3}\,\tilde g_{ij}\, \widetilde K\\
\widetilde A^{ij} &=& \psi^{-10} A^{ij}\\
\widetilde K &=& K\\
\label{eq:Adecomp}
 A^{ij} &=&  A^{ij}_* + (\Long W)^{ij}\,,
\end{eqnarray}
where $\tilde A^i\,_i = A^i\,_i= 0$ and $\nabla_iA^{ij}_* = 0$ with
$\nabla_i$ covariant differentiation with respect to the conformal
metric $g_{ij}$. In the tensorial decomposition of $A^{ij}$ given by
Eq.~(\ref{eq:Adecomp}), $A^{ij}_*$ gives the transverse part of
$A^{ij}$, with the longitudinal part given by
\begin{equation}
\label{eq:long}
(\Long W)_{ij} \equiv 2\,\nabla_{(i}\,W_{j)} - \frac{2}{3}\,g_{ij}\,\nabla_k\,W^k\,.
\end{equation}
With the transformations Eqs.~(\ref{eq:gconf}-\ref{eq:Adecomp}), the
constraint Eqs.~(\ref{hamcons}) and (\ref{momcons}) become:
\begin{eqnarray}
  &&8\,\Delta\psi - \psi\, R 
  - \frac{2}{3}\psi^5 K^2 + \psi^{-7} A_{ij} A^{ij}  = -16\,\pi\psi^5\tilde\rho \label{eq:hc}\\
  &&(\Delta_LW)^i - \frac{2}{3}\,\psi^{-6}\,\nabla^i\, K = 8\,\pi\psi^{10}\tilde j^i\,,\label{eq:mc}
\end{eqnarray}
with $R$ the Ricci scalar associated with the conformal 3-metric
$g_{ij}$ and $(\Delta_LW)^i\equiv \nabla_j (\Long W)^{ij}$.

At this point, we introduce the assumptions of conformal
flatness $g_{ij}=\eta_{ij}$ and vanishing of both $K$ and
$A^{ij}_*$. These assumptions exhaust the eight freely specifiable
conditions at our disposal on \elpar{;} five are in $g_{ij}$, one in
$K$ and two in $A^{ij}_*$. The constraints then take the form:
\begin{eqnarray}
&&\Delta\psi  +\frac{1}{8}\,\psi^{-7}(\Long W)^2 = -2\,\pi\psi^5\tilde\rho\label{ctt:hc}\\
&&(\Delta_L W)^i = 8\,\pi\psi^{10}\tilde j^i\,,\label{ctt:mc}
\end{eqnarray}
where $(\Long W)^2 \equiv  (\Long W)^{ij} (\Long W)_{ij}$.
In the absence of matter sources, or if one sets $j^i = \psi^{10}\tilde j^i$,
the constraints Eqs.~(\ref{ctt:hc}) and (\ref{ctt:mc}) decouple.
That is, one can solve first Eq.~(\ref{ctt:mc}) for $W^i$
and use this solution to solve Eq.~(\ref{ctt:hc}) for $\psi$.

% We construct initial data using the puncture
% approach~\cite{Anninos:1995vf}, which requires specifying the
% coordinate locations and momenta for the two \bh{s}.  For ``circular''
% orbits, we follow~\cite{Husa:2007rh}.

Following Ref.~\cite{2004PhRvD..69l4029F}, with the help of the
momentum constraint Eq.~(\ref{ctt:mc}), we notice that
\begin{eqnarray}
\label{eq:lw2}
(\Long W)^2 &=& 2\,(\Long W)^{ij}\nabla_iW_j\nonumber\\
            &=& 2\,\nabla_i[(\Long W)^{ij}W_j] - 2\,W_j\nabla_i(\Long W)^{ij}\nonumber\\
            &=& 2\,\nabla_i[(\Long W)^{ij}W_j] - 16\,\pi\,\psi^{10}W_j\tilde j^j\,.
\end{eqnarray}
Substitution of Eq.~(\ref{eq:lw2}) into the Hamiltonian constraint
Eq.~(\ref{ctt:hc}) yields
\begin{equation}
\Delta\psi  +\frac{1}{4}\,\psi^{-7}\nabla_i[(\Long W)^{ij}W_j] 
= -2\,\pi[\psi^5\tilde\rho-\psi^{3}W_i\tilde j^i]\,.\label{ctt:hc2}
\end{equation}

We address now the matter sources. The stress-energy tensor for a set
of non-interacting point-like particles with rest mass ${\cal M}_A$, 4-velocity
$U_A^a$, and comoving number density ${\cal N}_A$ is given by
\begin{equation}
\label{eq:tmunu}
T^{ab} = \sum_A {\cal M}_A\,{\cal N}_A\,U_A^a\,U_A^b\,,
\end{equation}
where the sum is understood to run over all the particles. For each
particle $A$ located at $x^i_A$, the comoving number density is
given by a $\delta$-function as
\begin{eqnarray}
{\cal N}_A &=& \int \frac{1}{\sqrt{-\,^{(4)}g}} \delta^4[x^a-x_A^a(\tau)]d\tau\nonumber\\
&=& \frac{1}{\alpha\,U^t_A\,\sqrt{\tilde g}}\delta^3[x^i-x_A^i(t)]\nonumber\\
&=& \frac{\delta_A}{\gamma_A\,\sqrt{\tilde g}}\,,
\end{eqnarray}
with $\,^{(4)}g$ the determinant of 4-dimensional spacetime metric,
$\delta_A \equiv \delta^3(x^i-x^i_A)$, $\alpha$ the lapse function,
$\gamma_A = \alpha\,U_A^t = - N_aU_A^a$, and $N^a$ the future-directed
unit normal to the hypersurface $\Sigma_t$. The stress-energy tensor
can then be rewritten as
\begin{equation}
\label{eq:tmunu2}
T^{ab} = \sum_A \frac{{\cal M}_A\,\delta_A}{\gamma_A\,\sqrt{\tilde g}}\,U_A^a\,U_A^b\,.
\end{equation}
Given Eq.~(\ref{eq:tmunu2}), the matter sources take the form:
\begin{eqnarray}
\label{eq:rhobh}
\tilde\rho &=& N_a\,N_b\,T^{ab}\nonumber\\
&=& \sum_A \frac{{\cal M}_A\,\gamma_A\,\delta_A}{\psi^6\,\sqrt{\eta}}\,,
\end{eqnarray}
and
\begin{eqnarray}
\label{eq:jbh}
\tilde j^a &=& -\perp^a_b\,N_c\,T^{bc}\nonumber\\
&=& \sum_A \frac{{\cal M}_A\,\perp^a_b\,U^b_A\,\delta_A}{\psi^6\,\sqrt{\eta}}\nonumber\\
&=& \sum_A \frac{P^a_A\,\delta_A}{\psi^{10}\,\sqrt{\eta}}\,,
\end{eqnarray}
where we have used $\sqrt{\tilde g} = \psi^6\,\sqrt{\eta}$, $g_{ab} =
\,^{(4)}g_{ab} + N_a\,N_b$ and $\perp^a_b =
\,^{(4)}g^{ac}g_{cb}$. In deriving Eq.~(\ref{eq:jbh}), we have also
introduced the spatial momentum vector $P_A^a \equiv {\cal
  M}_A\,\psi^4\,\perp^a_b\,U^b_A$. The vector $P^a$ is related to
the spatial part of the 4-momentum $p^a = {\cal M} U^a$ of the
point-like particles by $P^a =
\psi^4\,\perp^a_b\,p^b$. Substitution of the source
Eqs.~(\ref{eq:rhobh}) and (\ref{eq:jbh}) into Eqs.~(\ref{ctt:mc}) and
(\ref{ctt:hc2}) yields
\begin{eqnarray}
\label{ctt:hc3}
&&\Delta\psi  +\frac{1}{4\,\psi^7}\nabla_i[(\Long W)^{ij}W_j] 
= -2\,\pi\sum_A\frac{m_A\,\delta_A}{\sqrt{\eta}} \\
\label{ctt:mc3}
&&(\Delta_L W)^i = 8\,\pi \sum_A \frac{P^i_A\,\delta_A}{\sqrt{\eta}}\,,
 \end{eqnarray}
where 
\begin{equation}
\label{eq:mbare}
m_A = \frac{{\cal M}_A\,\gamma_A}{\psi} -\frac{W_iP_A^i}{\psi^7}\,.
\end{equation}

\citet{Bowen:1980yu} found a solution to the momentum constraint as
given by Eq.~(\ref{ctt:mc3}).  The solution represent \bh{s} with
linear momentum $P^i_A$ and is explicitly given as
\begin{equation}
\label{wlin}
W^i = -\sum_A\left .\frac{1}{4\,r} (7\,P^i + n^i\,n_j\,P^j)\right |_A 
\end{equation}
with $n^i$ the unit normal of constant $r$ spheres in flat space.
In terms of Eq.~(\ref{wlin}), $(\Long W)^{ij}$ takes the form:
\begin{equation}
\label{alin}
(\Long W)^{ij} = \sum_A\frac{3}{2\,r^2}\left[ 2\,P^{(i}\,n^{j)} 
- (\eta^{ij} - n^i\,n^j)\,n_k\,P^k\right]_A 
\end{equation}
In Eqs.~(\ref{wlin}) and (\ref{alin}), $r_A = ||x^i-x^i_A||$, $n^i_A =
(x^i-x^i_A)/r_A$ with $x^i_A$ the coordinate location of \bh{$_A$}.
It can be shown that the total ADM linear momentum is $P^i
= \sum_AP^i_A$.

We now turn our attention to the Hamiltonian constraint
Eq.~(\ref{ctt:hc3}).  As pointed out in
Ref.~\cite{2004PhRvD..69l4029F}, the term $\psi^7\nabla_i[(\Long
W)^{ij}W_j]$ in Eq.~(\ref{ctt:hc3}) is a ``flesh'' term that provides
the field between the particles and has the following contribution to
the Hamiltonian:
\begin{equation}
\nonumber
\int \frac{1}{\psi^7}\nabla_i[(\Long W)^{ij}W_j]d^3x  =
-7\int\frac{1}{\psi^8}(\Long W)^{ij}W_j\nabla_i\psi\,d^3x \,.
\end{equation}
The only approximation that goes into defining the skeleton
initial data is to neglect the contribution from this term. With this
approximation, the Hamiltonian constraint Eq.~(\ref{ctt:hc3}) reads:
\begin{equation}
\label{hc:skeleton}
\Delta\psi = -2\,\pi\sum_A\frac{m_A\,\delta_A}{\sqrt{\eta}}\,
\end{equation}
with $m_A$ given by Eq.~(\ref{eq:mbare}). Notice that $m_A$ is
singular at $x^i = x^i_A$ because $\psi$ and $W^i$ are singular at
$x^i_A$. Following Ref.~\cite{2004PhRvD..69l4029F}, we solve
Eq.~(\ref{hc:skeleton}) by means of Hadamard's ``partie finie''
procedure~\cite{Jaranowski:1999zu}; that is,
\begin{eqnarray}
\label{eq:psi}
  \psi &=&1 -4\,\pi\Delta^{-1}\left(\sum_A \frac{m_A(x^i)\,\delta_A}{2\,\sqrt{\eta}}\right)\nonumber\\
  &=& 1-4\,\pi\Delta^{-1}\left(\sum_A \frac{m^{\mathrm{(reg)}}_A(x_A^i)\,\delta_A}{2\,\sqrt{\eta}}\right)\nonumber\\
  &=& 1-4\,\pi\sum_A \frac{m^{\mathrm{(reg)}}_A(x_A^i)}{2}\,\Delta^{-1}\frac{\delta_A}{\sqrt{\eta}}\nonumber\\
  &=&1+\sum_A\frac{m^{\mathrm{(reg)}}_A}{2\,r_A}\,,
\end{eqnarray}
where
\begin{eqnarray}
  m^{\mathrm{(reg)}}_A &\equiv& \frac{{\cal M}_A\,\gamma_A}{\Phi_A} -\frac{W^A_iP_A^i}{\Phi_A^7}\\
  \Phi_A  &=& 1 + \sum_{B\ne A} \frac{m^{\mathrm{(reg)}}_B}{2 r_{AB}}\\
  \gamma_A &=& \left[1+\frac{P^iP_i}{{\cal M}^2\Phi^4}\right]^{1/2}_A\\
  W^A_i\,P^i_A &=& \sum_{B\ne A} \left(\frac{-1}{4\,r_{AB}}\right)[7\,P^i_B\,P_i^A \nonumber\\
  &-& (n^{AB}_i\,P^i_A)(n^{AB}_i\,P^i_B)]  \,, \label{eq:wp}
\end{eqnarray}
with $r_{AB} = ||x_A^i-x^i_B||$ and $n^i_{AB} =
(x^i_A-x^i_B)/r_{AB}$. The parameter $m_A$ is commonly known as the
\emph{bare} mass of the \bh{.}  On the other hand, ${\cal M}$ is known
as the irreducible mass of the \bh{.} $\Phi_A$ is the regularized
value of $\psi(x^i_A)$. In summary, the skeleton
initial data \elpar{} is then given by $\tilde g_{ij} = \psi^4\eta_{ij}$
and $\widetilde K_{ij} = \psi^{-2}(\Long W)_{ij}$ with $\psi$ given by
Eq.~(\ref{eq:psi}) and $(\Long W)^{ij}$ given by Eq.~(\ref{alin}).

For comparison, the exact or constraint-satisfying puncture initial
data method~\cite{Brandt97b} consists also of $\tilde g_{ij} = \psi^4\eta_{ij}$ and
$\widetilde K_{ij} = \psi^{-2}(\Long W)_{ij}$ with $(\Long W)^{ij}$
given by Eq.~(\ref{alin}), but in this case
\begin{equation}
\label{eq:psi_punc}
\psi =1+\sum_A\frac{m_A}{2\,r_A} + u\,,
\end{equation}
with $u$ a regular solution to
\begin{equation}
\label{eq:uu_punc}
\Delta u  +\frac{1}{8\,\psi^7}(\Long W)^2 = 0\,.
\end{equation}

%%%%%%%%%%%%%%%%%%%%%%%%%%%%%%%%%%%%%%%%%%%%%%%%%%%%%%%%%%%%%%%%%%%%%%%%%%%
\section{Quasi-circular Initial Data}
\label{sec:quasi}

We restrict our attention to initial data configurations representing
two equal mass (${\cal M}_1 = {\cal M}_2 \equiv {\cal M}$,
$m^{\mathrm{(reg)}}_1=m^{\mathrm{(reg)}}_2\equiv m$), non-spinning \bh{s} in
quasi-circular orbits. That is $P^i_1 = - P^i_2 \equiv P^i$, $r_{12} =
||x_1^i-x^i_2|| \equiv d$, and $n_i^{12}P^i = 0$. Under these
assumptions:
\begin{equation}
\label{eq:psi_final}
\psi =1+\frac{m}{2\,r_1}+\frac{m}{2\,r_2}
\end{equation}
where
\begin{eqnarray}
\label{eq:mbare2}
m &=& \frac{{\cal M}\,\gamma}{\Phi} -\frac{7}{4}\frac{P^2}{d\,\Phi^7}\\
\Phi  &=& 1 + \frac{m}{2\,d}\\
\gamma &=& \left[1+\frac{P^2}{{\cal M}^2\Phi^4}\right]^{1/2}\,.
\end{eqnarray} 
While deriving Eq.~(\ref{eq:mbare2}), we used that for circular orbits
$W_iP^i = 7\,P^2/(4\,d)$ with $P^2 = P^iP_i = P^iP^j\eta_{ij}$ as can be
seen from Eq.~(\ref{eq:wp}).

We focus now on the differences between the constraint-satisfying and
skeleton initial data for quasi-circular sequences using the
effective potential method~\cite{Cook:2000vr}.  The general idea of
this method is to find configurations that satisfy the condition:
\begin{equation}
\label{eq:epot}
\left . \frac{\partial E_b}{\partial l} \right |_{M,J} = 0\,,
\end{equation}
with $E_b = E-M$ the binding energy of the system. The distance $l$ is
a measure of the proper separation between the \bh{s} (e.g. horizon to
horizon), and $M = 2\,{\cal M}$ is the sum of the irreducible masses.
The quantities $E$ and $J$ are respectively the total ADM mass and
angular momentum of the system~\cite{Bowen80}, which can be computed
from:
\begin{eqnarray}
\label{eq:Madm}
E &=& -\frac{1}{2\,\pi}\oint_\infty \nabla_i\psi\,d^2S^i\\
\label{eq:Jadm}
J_i &=& \frac{\epsilon_{ijk}}{8\,\pi}\oint_\infty x^j\widetilde K^{kl}\,d^2S_l\,.
\end{eqnarray}
It is not too difficult to show from Eq.~(\ref{eq:Jadm}) that, given
$\widetilde K_{ij} = \psi^{-2}(\Long W)_{ij}$, the ADM angular
momentum for binaries initially in quasi-circular orbits is $J =
d\,P$. On the other hand, with $\psi$ given by
Eq.~(\ref{eq:psi_final}) the total ADM mass from Eq.~(\ref{eq:Madm})
is given by the sum of the bare masses of the \bh{s}, namely $E =
2\,m$; thus, the binding energy becomes $E_b = 2\,m - 2\,{\cal M}$.
The bare masses for the skeleton initial data are obtained by solving 
the implicit Eq.~(\ref{eq:mbare2}) using a Newton-Raphson method.

\begin{figure}
%scale=0.5
\includegraphics[width=80mm]{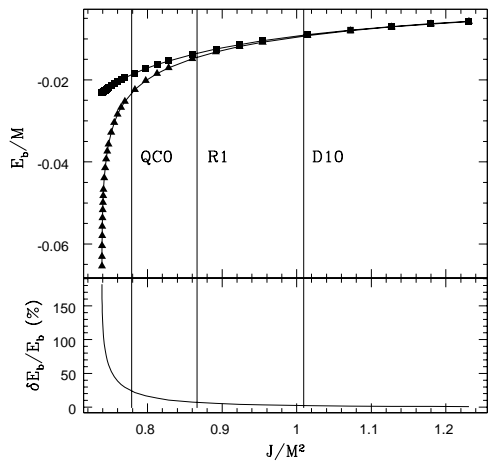}
\caption{Comparison of the binding energy $E_b$ as a function of the
  total ADM angular momentum $J$ between the initial data from
  \citet{Tichy:2003qi} (squares) and the skeleton initial data
  (triangles)}
\label{fig:Ebind}
\end{figure}

Figure~\ref{fig:Ebind} (top panel) shows the comparison of the binding
energy $E_b$ as a function of the total ADM angular momentum $J$
between the constraint-satisfying initial data from
\citet{Tichy:2003qi} (squares) and the skeleton initial data in
this work (triangles). The lower panel in Fig.~\ref{fig:Ebind} shows
the corresponding $\%$ relative difference between both results. Not
surprisingly, as the binary separation increases (i.e. larger angular
momentum), the differences diminish. For reference, the vertical lines
denote the angular momentum for typical data sets considered in the
literature: QC0 in Ref.~\cite{Campanelli:2005dd}, R1 in
Ref.~\cite{Baker:2006yw} and D10 in Ref.~\cite{Tichy:2003qi}. The
differences in binding energy between the skeleton and the
constraint-satisfying initial data are $\sim 20\,\%$ for QC0, $\sim
6\,\%$ for R1 and $\sim 2\,\%$ for D10.

Table~\ref{tbl:ID} provides the parameters of the initial
configurations for both the skeleton and constraint-satisfying
data sets. The cases of exact or constraint-satisfying initial 
data are labeled with the letter ``e'' and the corresponding
skeleton or approximate case with the letter ``a''.

\begingroup
\squeezetable
\begin{table}
  \begin{center}
\begin{ruledtabular}
\begin{tabular}{c|cccccc}
  Run & $d/M$ & $P/M$ & $m/M$ & ${\cal M}/M$ & $E/M$ & $J/M^2$ \\
  \hline %------------------------------------
 QC0e &   2.337 &  0.33320 & 0.45300   & 0.519071 & 1.0195 & 0.7787 \\ 
 QC0a &   2.337 &  0.33320 & 0.48950   & 0.519071 & 0.9790 & 0.7787 \\
 R1e  &   6.514 &  0.13300 & 0.48300   & 0.505085 & 0.9957 & 0.8664 \\
 R1a  &   6.514 &  0.13300 & 0.49717   & 0.505085 & 0.9943 & 0.8664 \\
 D10e &  10.00  &  0.09543 & 0.48595   & 0.500000 & 0.9895 & 0.9530 \\
 D10a &  10.00  &  0.09543 & 0.49458   & 0.500000 & 0.9891 & 0.9530 
\end{tabular}
\end{ruledtabular}
\end{center}
\caption{\emph{Initial data parameters:} The punctures have bare masses
  $m$, linear momenta $\pm P$ and are separated by a distance $d$.  The 
  irreducible mass of each \bh{} from $m^{\mathrm{(reg)}}$ is ${\cal M}$. 
  The ADM masses and angular momenta of the spacetimes are given 
  respectively by $E$ and $J$}.
\label{tbl:ID}
\end{table}
\endgroup

\begin{figure}
%scale=0.5
\includegraphics[width=80mm]{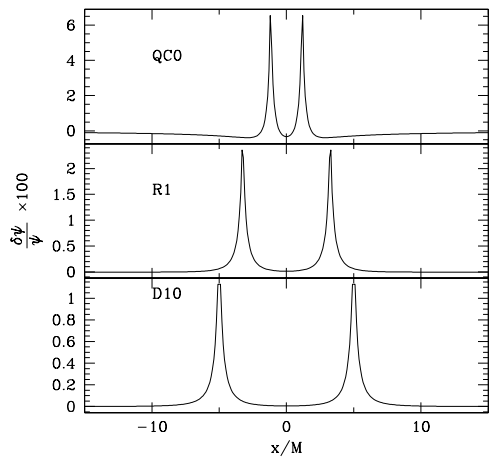}
\caption{The relative difference in the conformal factor $\psi$
  between the skeleton initial data and the corresponding
  constraint-satisfying data along the $x$-axis joining the
  punctures for the three cases labeled in Fig.~\ref{fig:Ebind}.}
\label{fig:conformal}
\end{figure}

As mentioned before, the only fundamental difference between the two
initial data sets is in the conformal factor $\psi$. For the
constraint-satisfying data set $\psi$ is computed from
Eq.~(\ref{eq:psi_punc}) by solving the Hamiltonian constraint in the
form given by Eq.~(\ref{eq:uu_punc}) and for the skeleton the
conformal factor $\psi$ is constructed using Eq.~(\ref{eq:psi}).  In
Fig.~\ref{fig:conformal}, we show the relative difference
$\delta\psi/\psi = (\psi_\mathrm{a}-\psi_\mathrm{e})/\psi_\mathrm{e}$
from the two data along the axis joining the punctures ($x$-axis) for
the D10, R1 and QC0 cases. Notice the large differences in the
immediate vicinity of the punctures. In the next section, we will
investigate how these differences translate into constraint
violations.

%%%%%%%%%%%%%%%%%%%%%%%%%%%%%%%%%%%%%%%%%%%%%%%%%%%%%%%%%%%%%%%%%%%%%%%%%%%%
\section{Hamiltonian Constraint Violations}
\label{sec:negcvs}

For the remainder of the paper we will concentrate our attention on
the D10 case: a situation in which the \bh{s} are not too close to the
merger and with an initial separation that permits a reasonable overlap
with the post-Newtonian regime~\cite{Baker:2006ha,Hannam:2007ik}.  It
is important to point out that the numerical data D10e, although
called exact, also violate the constraints initially.  The violations
in the exact initial data, however, are a consequence of numerical
errors which can be made arbitrarily small in the limit to the
continuum. On the other hand, the constraint violations in the
skeleton data are independent of the resolution of the computational
grid.

In order to understand the nature of the constraint violations in the
skeleton initial data and in particular their dynamics in the course
of the evolution, we take the point of view that the violations
introduce ``spurious'' sources $\tilde{\rho}$ and
$\tilde{j}^i$ in Eq.~(\ref{hamcons}) and (\ref{momcons}),
respectively. Notice that initially we do not have a ``spurious''
momentum density $\tilde{j}^i$ because the skeleton initial data by
construction are an exact solution to the momentum constraint.  It is
important to keep in mind that one should not assign physical
properties to $\tilde{\rho}$ and $\tilde{j}^i$. They are only used to
quantify constraint violations. In particular, the violations
$\tilde{\rho}$ are not restricted to satisfy energy conditions and
thus are free to take negative values.

\begin{figure}
\includegraphics[width=80mm]{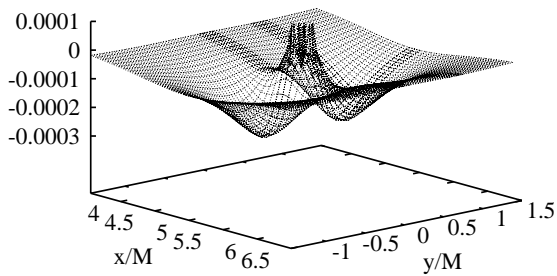}
\caption{Surface plot of $\tilde{\rho}$, as derived 
  from the Hamiltonian constraint violations, in the 
  $xy$-plane surrounding one puncture for the skeleton initial data, D10a.}
\label{fig:hamconst}
\end{figure}

\begin{figure}
\includegraphics[width=80mm]{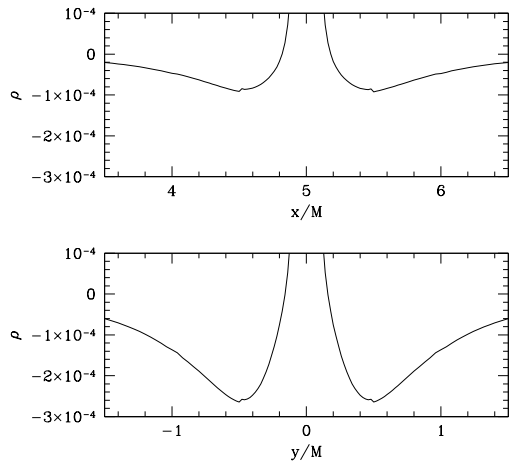}
\caption{ Sources $\tilde{\rho}$ corresponding
  to Fig.~\ref{fig:hamconst} along the $x$-axis joining the \bh{s}
  (top panel) and along the $y$-axis (bottom panel), the linear
  momentum direction.}
\label{fig:rho1d}
\end{figure}

Fig.~\ref{fig:hamconst} shows a surface plot of $\tilde{\rho}$ for the
\bbh{} skeleton initial data in the neighborhood of one of the
punctures.  Notice that the puncture seems to be embedded in a
``cloud'' or a pocket of negative $\tilde{\rho}$. Furthermore, the
cloud is more negative in the direction aligned with the linear
momentum of the puncture (in this case the $y$-axis). This effect is
more evident from Fig.~\ref{fig:rho1d} where we plot $\tilde{\rho}$ in
the top panel along the $x$-axis (the direction joining the \bh{s})
and in the bottom panel along the $y$-axis.  The glitches at the bottom
of the bottom of the constraint violation pockets are due to refinement 
boundaries.

%%%%%%%%%%%%%%%%%%%%%%%%%%%%%%%%%%%%%%%%%%%%%%%%%%%%%%%%%%%%%%%%%%%%%%%%%%%
\section{Skeleton Evolutions}
\label{sec:evolutions}

Given the initial data, we turn our attention now to evolutions. 
The evolution runs were done on a computational grid with 9 refinement
levels, the finest 5 levels containing $24^3$ gridpoints in radius and the
remaining 4 with $48^3$ gridpoints in radius. To check the dependence of the
results with resolution, we considered grid spacings at the finest
level of $M/38.4$, $M/44.8$ and $M/51.2$.  The results presented here 
were done at the resolution of $M/51.2$.

\begin{figure}
\includegraphics[width=80mm]{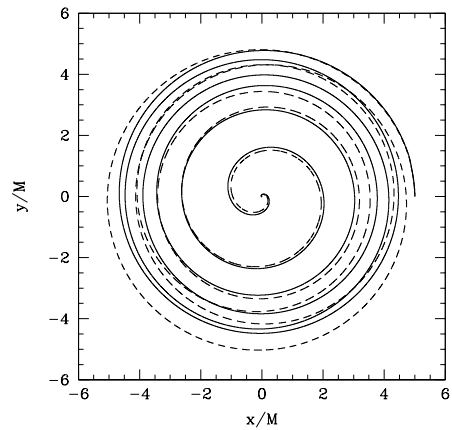}
\caption{Trajectory of one of the \bh{s} from the skeleton
  initial data (dashed line) as well as its constraint-satisfying
  counterpart (solid line).}
\label{fig:CSCVtrajs}
\end{figure}

Fig.~\ref{fig:CSCVtrajs} shows the trajectory of one of the \bh{s}
from the skeleton initial data (dashed line) as well as its
constraint-satisfying counterpart (solid line). Both trajectories are
very close to each other during the first quarter orbit. Beyond that
point, the \bh{} from the skeleton initial data follows an eccentric
orbit.  Finally, near merger or at the plunge, the trajectories once
again lie very closely together.

\begin{figure}
\includegraphics[width=80mm]{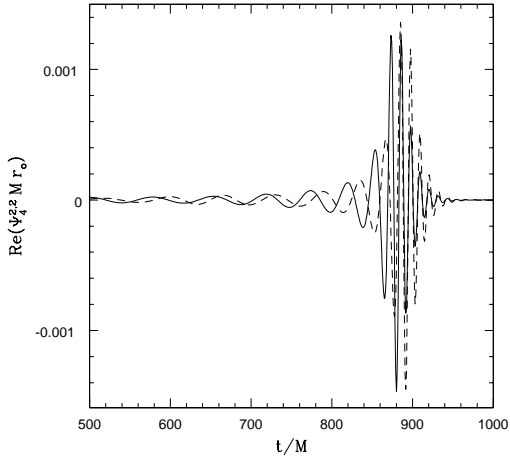}
\caption{Real parts of the waveform, $r_o \Psi^{2,2}_4\,M$, extracted
  at $r_o=50\,M$ for both the skeleton (dashed line) and
  the constraint-satisfying (solid line) initial data.}
\label{fig:CSCVwaves}
\end{figure}

\begin{figure}
\includegraphics[width=80mm]{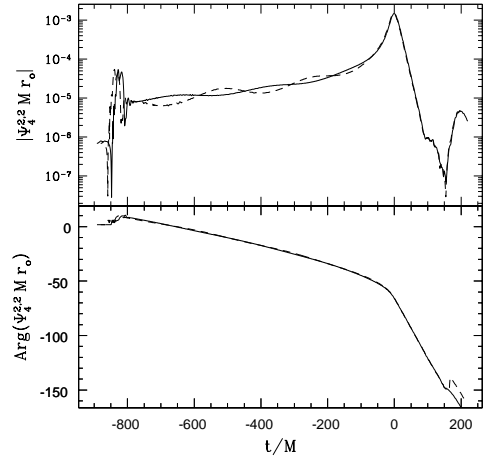}
\caption{Amplitude (top panel) and phase (bottom panel) of the
  waveforms $r_o \Psi^{2,2}_4\,M$ in Fig.~\ref{fig:CSCVwaves},
  skeleton data (dashed line) and constraint-satisfying data
  (solid line). The time axis has been shifted by $10\,M$ to align the
  point at which the amplitudes reach their maximum values.}
\label{fig:CSCVamps}
\end{figure}

In Fig.~\ref{fig:CSCVwaves}, we compare the waveforms of the skeleton
initial data with its constraint-satisfying counterpart as detected at
$50\,M$. The presence of a phase shift between the two waveforms is
evident. The constraint-satisfying initial data evolution reaches the
merger approximately $10\,M$ before the skeleton initial data
evolution. This difference remains within $1\,M$ of this between different
resolutions.  Another difference in the two evolutions
is in the inspiral.  As mentioned before, the skeleton data yields a
larger eccentricity in the inspiral.  This can be clearly observed
from Fig.~\ref{fig:CSCVamps} where the same comparison as in
Fig.~\ref{fig:CSCVwaves} is shown but in terms of the amplitude (top
panel) and phase (bottom panel). Here we have applied a time shift of
$10\,M$ to align the point at which the waveforms reach their maximum
values. The inspiral and plunge of the binary is before the ``knee''
in the phase or the maximum in the amplitude. On the other hand the
quasi-normal ringing of the final \bh{} takes place after the
knee in the phase and the maximum in the amplitude. Notice that the
phases are practically identical for both cases. Furthermore,
both the post-knee phase and post-maximum amplitude are almost the
same for skeleton and constraint-satisfying evolutions, which is an
indication that the final \bh{s} are almost identical~\cite{Hinder:2007qu}. On the other
hand, the inspiral amplitudes in Fig.~\ref{fig:CSCVamps} clearly show
differences in the level of eccentricity as seen by the oscillations
in the amplitude.

From the waveforms, we have computed the energy $E_{\hbox{rad}}$ and
angular momentum $J_{\hbox{rad}}$ radiated. For the
constraint-satisfying initial data, we obtained $E_{\hbox{rad}} =
0.0354\,M$ and $J_{\hbox{rad}} = 0.3060\,M^2$ and for the skeleton
data $E_{\hbox{rad}} = 0.0359\,M$ and $J_{\hbox{rad}} = 0.3063\,M^2$,
which correspond to differences of $1.4\%$ and $0.1\%$ respectively.
These differences are consistent with differences in energy and
angular momentum in the initial data ($< 10^{-4}\%$).

\begin{figure}
\includegraphics[width=80mm]{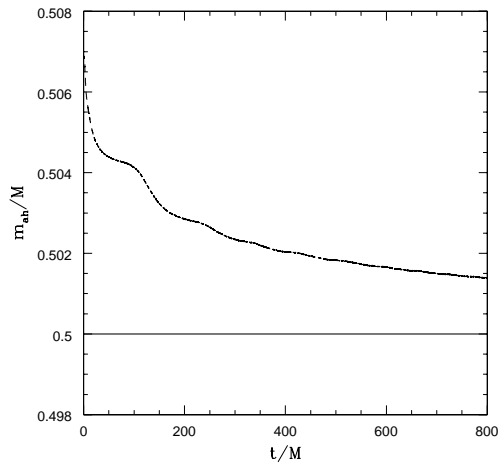}
\caption{The evolution of the \ahz{} mass of one of the \bh{s} shown
  for both the constraint-satisfying initial data D10e (solid line) 
  and its skeleton counterpart (dashed line). Errors in \ahz{} due 
  to grid spacing are of order $10^5\,M$.}
\label{fig:CSCVmoft}
\end{figure}

To better understand the change in trajectories and the corresponding
phase shift reflected in the waveforms (see Fig.~\ref{fig:CSCVwaves}),
we have tracked the evolution of the \ahz{} masses. The \ahz{} mass
for one of the \bh{s} is plotted in Fig.~\ref{fig:CSCVmoft} where the 
error due to grid spacing resolution is of order $10^-5\,M$.  While
the \ahz{} mass for the constraint-satisfying evolution stays
relatively constant (solid line), the \ahz{} mass for the skeleton
evolution varies significantly (dashed line). In fact, the mass starts
more than 10\% higher than the constraint-satisfying value and
monotonically decreases.  Empirically, the \ahz{} masses decrease as
$1/t$ at late times. By fitting a polynomial in $1/t$ to the \ahz{} evolution at 
late times, we find the mass asymptotes to $0.501 \pm 0.001
M$, within 0.2\% of the constraint-satisfying initial \ahz{} mass.
However, the \bh{s} merge before the skeleton \ahz{} mass could reach
this asymptotic value.  The differences and evolution of the \ahz{} 
masses early in the evolution of the skeleton data are 
consistent with the picture of a binary whose masses and therefore
binding energy and dynamics are altered. 

%%%%%%%%%%%%%%%%%%%%%%%%%%%%%%%%%%%%%%%%%%%%%%%%%%%%%%%%%%%%%%%%%%%%%%%%%%%%
\section{Single Puncture Analysis}
\label{sec:single_punc}

As noted in Sec.~\ref{sec:quasi}, the Hamiltonian constraint
violations are negative in the vicinity of the punctures.  To better
understand the evolutions of the skeleton initial data, we consider a
test case where we evolve a single, non-spinning puncture and add
by hand negative constraint violations surrounding it. That is, we
solve the Hamiltonian constraint as if there were an additional matter
field $\tilde{\rho}$ present, namely
\begin{equation}
\Delta \psi = -2\,\pi\,\tilde\rho\,\psi^5\,.
\label{eq:single_punc}
\end{equation}
In order to guarantee the existence of a solution as discussed in
\cite{1979sgrr.conf...83Y,Choquet00}, if $\tilde{\rho}>0$, one needs
to re-scale the source according to the conformal rescaling
$\tilde{\rho} = \rho\, \psi_o^{-s}$, with $s>5$ . In our case,
however, we are mostly interested in $\tilde{\rho}<0$, which does not
require any rescaling for existence of a solution.

Following the procedure for multiple \bh{s}, see
Eq.~(\ref{eq:psi_punc}), we use the ansatz $\psi = \psi_o + u$, with
$\psi_o = 1+m/2r$ the solution to the homogeneous equation (i.e. the
single puncture solution). With this ansatz,
Eq.~(\ref{eq:single_punc}) becomes
\begin{equation}
\Delta u = -2\,\pi\,\rho\,(\psi_o+u)^n\,
\label{eq:single_punc2}
\end{equation}
where $n=-3$ for $\tilde{\rho}>0$ and $n=5$ for $\tilde{\rho}<0$.

For simplicity, we choose 
\begin{equation}
\rho= \psi_o^{m} F e^{-(r-r_o)^2/(2 \sigma^2)}
\end{equation} 
where $r_0$ is the position with respect to the puncture and $m=0$ for
$\tilde{\rho}>0$ and $m=-5$ for $\tilde{\rho}<0$. The factor
$\psi_o^{m}$ is necessary for regularity of the solution $u$ at the
puncture.  We also assume that the source $\tilde{\rho}$ does not have
initial momentum (i.e. $\tilde j^i = 0$); thus, the momentum
constraint remains satisfied as in the vacuum case.

Table~\ref{tbl:id_models} lists the results from the evolutions for
$r_o = \sigma = 1\,M$. Notice that model $F_1$ has a positive source
(i.e. $F>0$) and the other two have negative Hamiltonian constraint
violations.  The effect of the source $\rho$ is evident in the ADM
mass ($E$) and initial \ahz{} mass (${\cal M}^i_{AH}$). For the
positive source, the masses are larger than the puncture mass in
vacuum, $1\,M$, and smaller for the negative sources.  Also in
Table~\ref{tbl:id_models}, we include $E_{\rho} = E-{\cal M}^i_{AH}$,
which gives a measure of the extra energy content in the initial data
due to $\rho$.

\begingroup
\squeezetable
\begin{table}
  \begin{center}
\begin{ruledtabular}
\begin{tabular}{c|rrrrr}
  Model & $F/M^2$ & $E/M$ & ${\cal M}^i_{AH}/M$ & ${\cal M}^f_{AH}/M$
  & $E_{\rho}/M$\\
\\
  \hline %------------------------------------
   $F_1$ &   0.001 & 1.0046 & 1.0012 & 1.0041 & 0.0034\\
   $F_2$ &  -0.001 & 0.9902 & 0.9973 & 0.9911 & -0.0071\\
   $F_3$ &  -0.010  & 0.9102 & 0.9858 & 0.9183 & -0.0756
\end{tabular}
\end{ruledtabular}
\end{center}
\caption{\emph{Models:}  
  Results of evolutions a single puncture in the presence of a
  Gaussian source
  $\rho$ with $r_o = \sigma = 1\,M$ and amplitude $F$. 
  The initial \ahz{} mass and ADM energy are ${\cal M}^i_{AH}$ 
  and $E$ respectively.  The \ahz{} mass 
  at the end of the simulation is given by ${\cal M}^f_{AH}/M$, and
  $E_{\rho} = E-{\cal M}^i_{AH}$.}
\label{tbl:id_models}
\end{table}
\endgroup

\begin{figure}
\includegraphics[width=80mm]{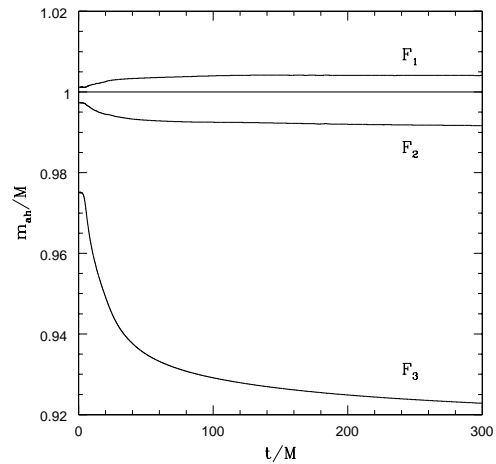}
\caption{Evolution of the \ahz{} mass for the models described in
  Table~\ref{tbl:id_models}.  The error between resolutions for F3 was
  of order $10^{-5}$ over the course of the evolution.}
\label{fig:model_mah}
\end{figure}

We evolved the models in Table~\ref{tbl:id_models} for $300\,M$.
Fig.~\ref{fig:model_mah} shows how the \ahz{} mass evolves during the
evolution.  We have evolved the model $F_3$ at different resolutions
and estimated the \ahz{} masses to have an approximate relative error 
of $\sim 0.002\%$.  We observed that at late times the \ahz{} mass evolves as
${\cal M}^f_{AH} + C/t$. The values reported in
Table~\ref{tbl:id_models} are those extrapolated to $t \rightarrow
\infty$.

The evolutions of the single puncture models clearly demonstrate that
depending on the signature of $\rho$, the mass of the \bh{,} as
measured from the \ahz{,} will increase or decrease. That is, over the
course of the evolution, the \ahz{} masses evolve to approach the ADM
energy, decreasing for a negative $\rho$ and increasing for 
positive $\rho$.  In other words, the source $\tilde{\rho}$ initially
hovering near a puncture will fall into the \bh{}, increasing or decreasing 
its mass as the system becomes stationary depending on the sign of $\tilde{\rho}$. 
The extent to which the final
mass of the \bh{} approaches the total ADM energy depends on how much
of the density $\rho$ is ``accreted'' by the \bh{.}  Since in our
case, we do not impose the restriction of positivity on $\rho$, the
\bh{} is free to decrease its mass. Notice also that the final \ahz{}
mass does not satisfy the condition ${\cal M}^f_{AH} = {\cal M}^i_{AH}
+ E_{\rho}$, which means that a fraction ($<1\%$ in our cases) of 
$E_{\rho}$ mass is radiated away. The choice of centering the Gaussian 
at $r_o = 1\,M$ was aimed at favoring the amount $\tilde{\rho}$ 
accreted by the \bh{.}

\begin{figure}
\includegraphics[width=80mm]{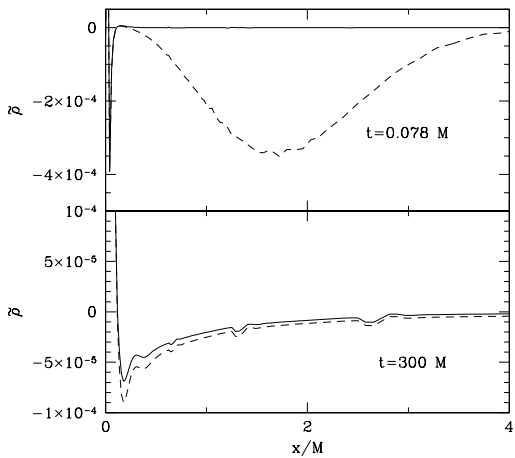}
\caption{Hamiltonian constraint violation $\tilde\rho$ near the
  beginning of the simulation at $t=0.078\,M$ (top panel) and at the
  end, $t=300\,M$, of the simulation (bottom panel). Solid
  lines represents the constraint-satisfying case and dashed lines the
  $F_3$ model. The constraint violations still present at late times are
  due to discretization around the punctures.}
\label{fig:ham_single}
\end{figure}

\begin{figure}
\includegraphics[width=80mm]{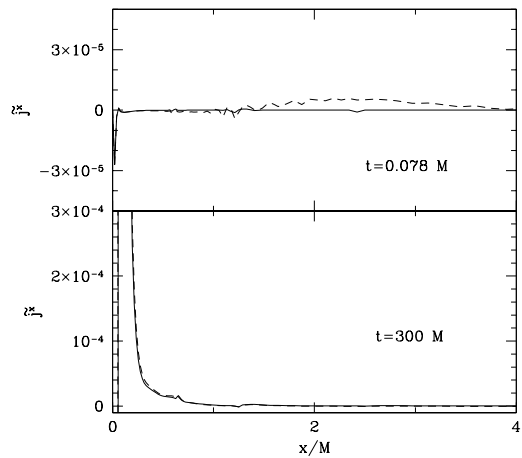}
  \caption{Same as in Fig.~\ref{fig:ham_single} but for the momentum
    constraint violation $\tilde j^x$. The constraint violations still
    present at late times are due to discretization around the puncture.}
\label{fig:mom_single}
\end{figure}

Figure~\ref{fig:ham_single} shows the Hamiltonian constraint violation
$\tilde\rho$ near the beginning of the simulation at $t=0.078\,M$ (top
panel) and at the end, $t=300\,M$, of the simulation (bottom
panel). Solid lines represent the constraint-satisfying case and
dashed lines the $F_3$ model. Fig.~\ref{fig:mom_single} shows the
corresponding results for the momentum constraint violations $\tilde
j^i$ along the $x$-axis.  By construction, initially there are only
Hamiltonian constraint violations in the $F_3$ model. However, it is
evident from the top panel in Fig.~\ref{fig:mom_single} that
constraint violations in the momentum constraint develop also very
early in the evolution. The growth of momentum constraint violations
proceed up to a time $t\sim 3\,M$. The subsequent dynamics of the
constraint violations consists of ingoing and outgoing waves. Because
of the proximity to the puncture, the outgoing waves are a little bit
weaker, with most of the constraint violations ``accreted'' by the
\bh{.} After approximately $t\sim50\,M$ of evolution, the $F_3$ model
relaxes to the configuration of the constraint-satisfying puncture 
and remains there as seen in the bottom panels in 
Figs.~\ref{fig:ham_single} and \ref{fig:mom_single}. The final constraint
violations in the system arise from numerical errors.

An important aspect to point out is that although the constraint-violating 
cases relax to a constraint-satisfying solution, the solutions that they 
relax to are not necessarily the same solutions as a single puncture in 
a vacuum spacetime.  The new solution satisfies the Einstein equations but
for a single puncture spacetime with a smaller mass.  A similar situation 
occurs in the binary case; the system relaxes to a binary solution, but 
this solution is different than the vacuum case.  The reason for this 
behavior is not currently understood.

%%%%%%%%%%%%%%%%%%%%%%%%%%%%%%%%%%%%%%%%%%%%%%%%%%%%%%%%%%%%%%%%%%%%%%%%%%%%
\section{Impact on Data Analysis}
\label{sec:dataanalysis}

Finally, we want to address the extent to which the waveforms from
evolutions of skeleton initial data may be of use in exploring
gravitational wave astronomy.  We will focus on computing the matches
between the skeleton and the constraint-satisfying waveforms. In
principle, the match would be between the detector output, $h_1$ and
the template, $h_2$.  Here $h_1$ is the waveform from the
constraint-satisfying evolution and $h_2$ from the skeleton initial
data evolution.  Specifically, we will compare the waveforms using the
minimax match given by \cite{finn-1992-46,Owen-96,damour-1998-57}.
\begin{equation}
\hbox{Match} \equiv  \max_{t_0}  \min_{\Phi_2} \max_{\Phi_1} \frac{ \langle h_1 |h_2
\rangle} {\sqrt{\langle h_1 | h_1 \rangle \langle h_2 | h_2 \rangle
}}\,,
\label{eqn:minimax}
\end{equation}
where the inner product of two templates is defined by
\begin{equation}
\langle h_1 | h_2 \rangle = 4 \, \mbox{Re}\int_\fmin^\fmax \frac{\tilde
h_1(f) \tilde h^*_2(f)}{S_{h}(f)} \, df.
\end{equation} 
The match is maximized over the time of arrival of the signal, $t_0$,
and minimized/maximized over the initial phase, $\Phi_1$ and $\Phi_2$,
of the orbit when the signal/template enters the LIGO band.  The
variable $S_h(f)$ denotes the noise spectrum for which we use the
initial LIGO noise curve~\cite{LIGO:SRD}.  The domain
$[f_{\mathrm{min}},f_{\mathrm{max}}]$ is determined by the detector
bandwidth and the masses of our signal -- set such that the initial
frequency of the numerical waveform just enters the LIGO band.  We
have chosen to study the match for values of the total mass of the
\bbh{} system greater than $20M_\odot$ because of the limited number
of cycles that our waveforms include.  A more detailed description of
our calculation of the minimax match is given in \cite{vaishnav-2007}.

\begin{figure}
\includegraphics[width=80mm]{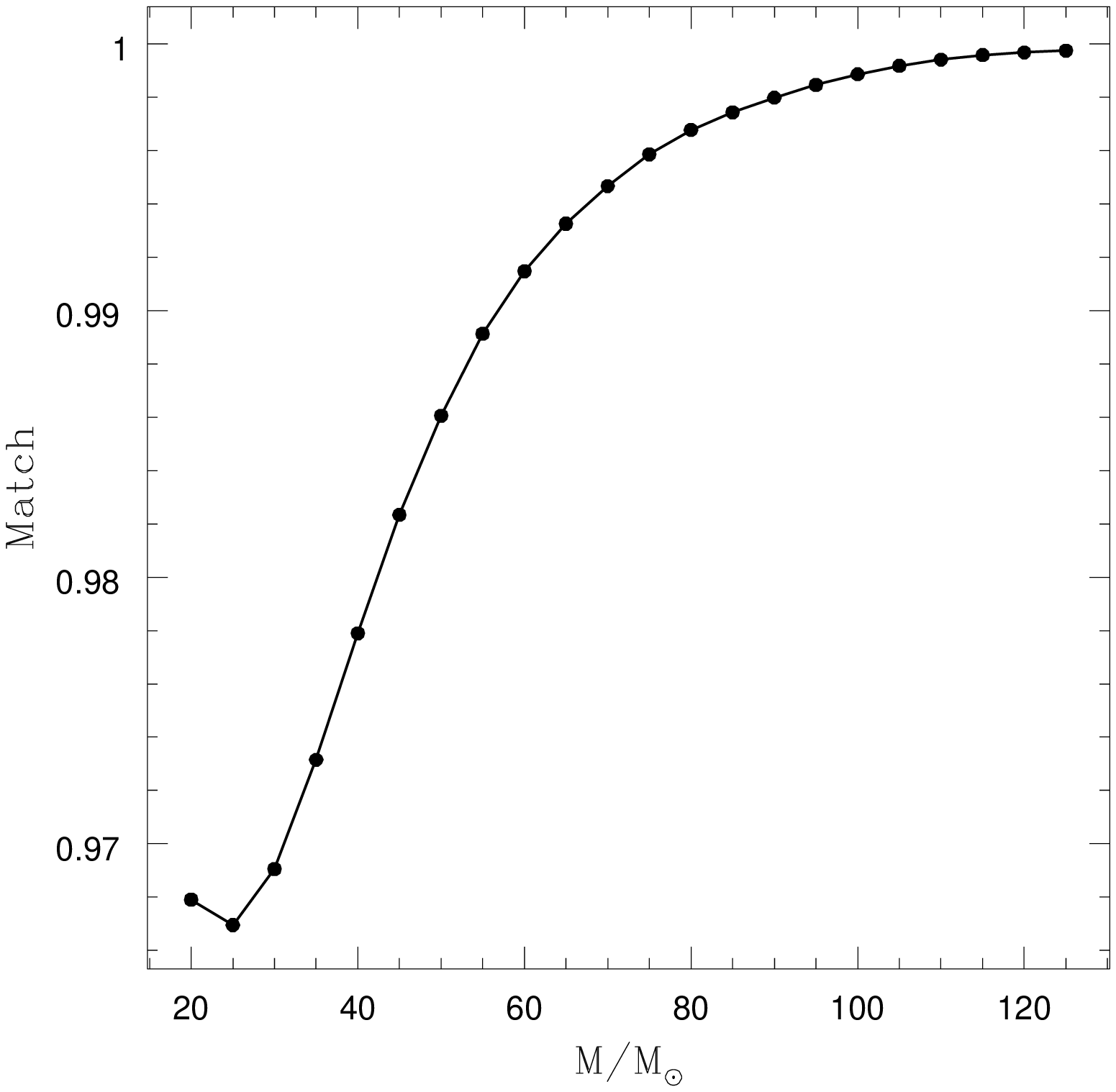}
\caption{Matches}
\label{fig:CSCVMismatch}
\end{figure}

The match between the constraint-satisfying and skeleton data versus
mass is plotted in Fig.~\ref{fig:CSCVMismatch}.  As the total mass
increases, the match between the waveforms increases, becoming $>0.99$
at $60M_\odot$.  At such large total mass, the signal is
dominated by the plunge and ring-down. Comparisons of the plunge and
ring-down show (see Fig.~\ref{fig:CSCVamps}) that the difference
between the skeleton and constraint-satisfying evolution are very
small. At masses lower than about $40M_\odot$, the match drops below
0.97 due to the difference in the binary dynamics prior to merger.
We note that our calculation of match did not maximize over the mass
of the two waveforms.  Maximizing over the mass would have diminished
the differences between the two waveforms.  

%%%%%%%%%%%%%%%%%%%%%%%%%%%%%%%%%%%%%%%%%%%%%%%%%%%%%%%%%%%%%%%%%%%%%%%%%%%%
\section{Conclusions}
\label{sec:conclusions}

We have carried out a study of the evolution of skeleton, puncture \bbh{}
initial data as proposed by \citet{2004PhRvD..69l4029F}.  We focused
on non-spinning punctures at initial separations of $10\,M$, where the
difference in binding energy with the constraint-satisfying initial
data is $< 2\%$.  We showed that during the inspiral the
skeleton data yields different dynamics; however, this difference
significantly diminishes as the binary enters the plunge, merger and
ring-down.

We tested the match between the constraint-satisfying and skeleton
data for a series of total masses between $20M_\odot$ and
$130M_\odot$.  Our results indicate that gravitational wave data
analysis would have some tolerance for constraint-violating data,
especially for those binaries in which the signal is plunge-merger
dominated, as is the case of high mass \bh{s}.  We conclude that
although the two systems were different, with one clearly violating
the Einstein equations, the differences were not enough to impact the
match statistics for the mass ranges we included in our study and for
the number of cycles present in our numerical waveforms.  If these
systems were evolved starting with a larger initial black-hole
separation, the constraint violations would be smaller; and,
therefore, the waveforms generated could be useful for detection over
the complete \bbh{} mass range for initial LIGO.  If, however, larger
constraint violations are present in the data that drive the early \bh{} mass'
lower, the differences may lead to errors in parameter estimation.
 
We also analyzed the impact of the Hamiltonian constraint
violations. We showed that the main feature of the skeleton data is
two packets of negative constraint violations in front of and behind the
\bh{}, along the direction of its momentum.  We conjectured that these
negative constraint violations acted as a source density that gets
absorbed by the \bh{s} during evolution. To test our conjecture, we
considered a model consisting of a single, non-rotating puncture in
which we artificially added a stationary Gaussian shell source that
mimics the Hamiltonian constraint violations in the skeleton data. The
evolutions of this single puncture model reproduce the decrease in the
mass of the \bh{} observed in the evolution of the skeleton data.

One remarkable aspect of our study is the ability of the BSSN
equations and moving puncture gauges to evolve stably data
away from the constraint surface. What is even more
remarkable is how the evolution brings the data back to the
Einstein constraint surface. We are currently investigating a broader
class of solutions with this property.

In summary, our numerical evolutions show that the skeleton initial
data proposed by \citet{2004PhRvD..69l4029F} embeds the \bh{s} in a
``cloud'' of negative constraint violations.  These constraint
violations act as a source field that when accreted by
the \bh{s} decreases their masses. The change in the masses modifies
the binding energy of the binary and thus affects its orbital dynamics
(e.g. adding eccentricity) but had little affect on the match of the two waveforms
for initial LIGO for high mass black holes. The observed effects will decrease as the
initial binary separation increases.

%%%%%%%%%%%%%%%%%%%%%%%%%%%%%%%%%%%%%%%%%%%%%%%%%%%%%%%%%%%%%%%%%%%%%%%%%%%%
\acknowledgments This work was supported in part by NSF grants
PHY-0354821, PHY-0653443, PHY-0244788, PHY-0555436, PHY-0801213 and PHY-0114375
(CGWP).  Computations were performed at NCSA and TACC under allocation
TG-PHY060013N.  The authors thank M.~Ansorg and E.~Schnetter for
contributions to the computational infrastructure.  \vfill
%%%%%%%%%%%%%%%%%%%%%%%%%%%%%%%%%%%%%%%%%%%%%%%%%%%%%%%%%%%%%%%%%%%%%%%%%%%%

\end{document}